\documentclass{elsarticle}
\usepackage[usenames,dvipsnames]{color}
\usepackage{graphicx}
\usepackage{amsmath}

\begin{document}
\journal{Physics Letters A}
\begin{frontmatter}
\title{Comment on 'Absolute negative mobility\\ in a one-dimensional overdamped system'} 
\author{J. Spiechowicz}
\author{M. Kostur}
\author{J. {\L}uczka\corref{cor1}}
\ead{jerzy.luczka@us.edu.pl}
\address{Institute of Physics, University of Silesia, 40-007 Katowice, Poland}
\address{Silesian Center for Education and Interdisciplinary Research, University of Silesia, \\41-500 Chorz{\'o}w, Poland}
\begin{abstract}
Recently Ru-Yin Chen {\it et al.} (Phys. Lett. A 379 (2015) 2169-2173) presented results on the absolute negative mobility (ANM) in a one-dimensional overdamped system and claimed that a new minimal model of ANM was proposed. We suggest that the authors introduced a mistake in their calculations. Then we perform a precise numerical simulation of the corresponding Langevin equation to show that the ANM phenomenon does not occur in the considered system.
\end{abstract}
\end{frontmatter}
%

In the paper \cite{chen2015} the authors consider overdamped motion of a particle in a one-dimensional \emph{symmetric} spatially periodic potential $V(x)$ under the influence of both  zero-mean trichotomous noise $\xi(t)$ \cite{mankin2000} and a constant force $F$. The corresponding dimensionless Langevin equation reads \cite{chen2015}
\begin{equation}
	\label{eq:model}
 	\dot{x} = -V'(x) + F + \xi(t),
\end{equation}
where the dot and prime denotes the differentiation with respect to time $t$ and the particle position $x(t) \equiv x$, respectively. The potential $V(x)=V(x+L)$ of the period $L = 1$ is assumed to be in a piecewise linear form, namely,  
\begin{equation}
    V(x) = \left\{ \begin{array}{ll} -(x-d)/d, & x \in (0,d)\, \textrm{mod} \, 1\\ (x-d)/(1-d), & x \in (d,1)\, \textrm{mod} \, 1.\\ \end{array} \right. 
\end{equation}
The parameter $d \in (0,1)$ controls its asymmetry. It is symmetric for $d = 1/2$ and this case was studied in Ref. \cite{chen2015}. Noise $\xi(t) = \{-a_0, 0, a_0\}$ is a three-state stationary Markov process. The stationary probabilities  are \cite{mankin2000}
\begin{figure}[t]
	\centering
	\includegraphics[width=0.45\linewidth]{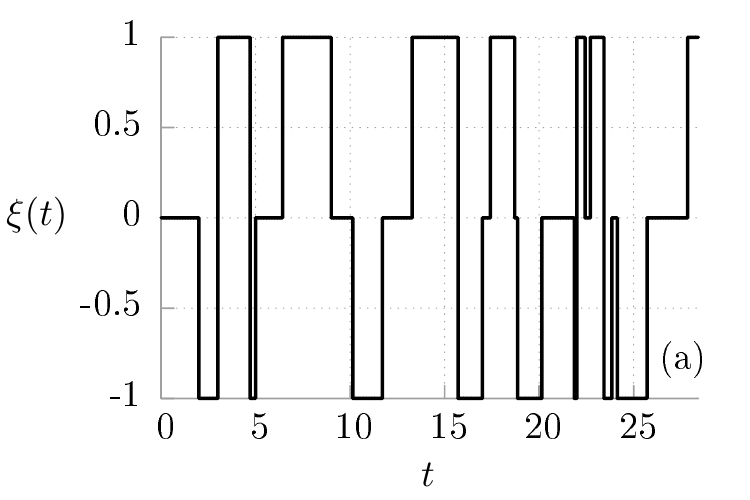}
	\includegraphics[width=0.45\linewidth]{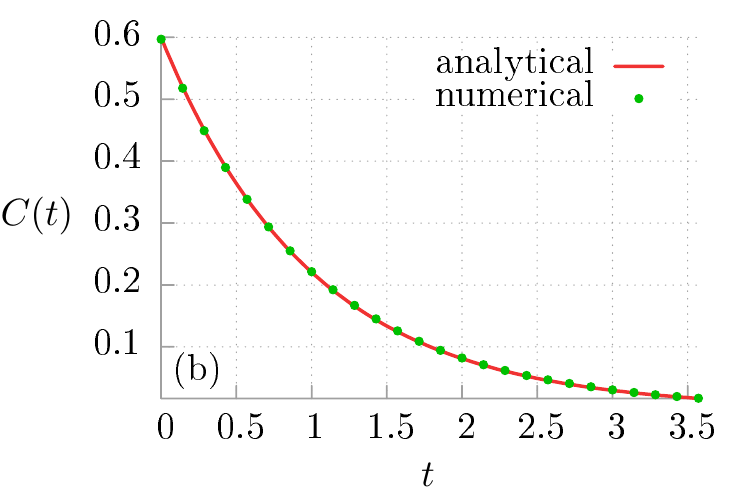}
	\caption{Illustrative realization of the trichotomous noise $\xi(t)$ and its correlation function $C(t)$ is presented in panel (a) and (b), respectively. In the latter we compare the direct numerical results with the analytical expression. Parameters are $q = 0.3$, $a_0 = 1$, $\nu = 1$.}
	\label{fig1}
\end{figure}	
\begin{equation}
	P_s(-a_0) = P_s(a_0) = q, \quad P_s(0) = 1 - 2q, \quad 0 < q < 1/2.
\end{equation}
The transition probabilities per unit time between these states are given by the relations
\begin{align}
	Pr(0 \to \pm a_0) &= Pr(-a_0 \to a_0) = Pr(a_0 \to -a_0) = q\nu, \nonumber\\
	Pr(\pm a_0 \to 0) &= (1 - 2q)\nu, \quad \nu > 0.
\end{align}
Consequently, the mean waiting times in the three states read  
\begin{equation}
	\tau_{a_0}=\tau_{-a_0} = \frac{1}{(1-q)\nu}, \quad \tau_{0} = \frac{1}{2q\nu}.
\end{equation}
The jumps form in time a Poisson process, i.e. the interval between successive transitions is exponentially distributed with the probability density characterized by the mean waiting time in the appropriate state. The mean value and the correlation function of trichotomous noise are
\begin{equation}
	\langle \xi(t) \rangle = 0, \quad C(t) = \langle \xi(s)\xi(s+t) \rangle = \langle \xi^2(s) \rangle e^{-\nu|t|} = 2qa_0^2 e^{-\nu|t|}.
\end{equation}
The switching rate $\nu$ is the reciprocal of the noise correlation time $\tau$, i.e. 
\begin{equation}
	\nu = 1/\tau.
\end{equation}
It is worth to note that trichotomous noise can be considered as a particular case of the kangaroo stochastic process \cite{kostur1999}.
\begin{figure}[t]
	\centering
	\includegraphics[width=0.45\linewidth]{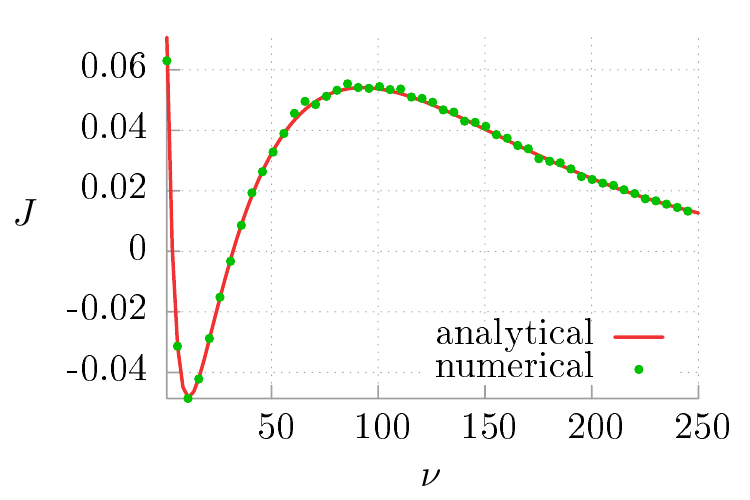}
	\caption{The probability current $J$ (which is equivalent  to the stationary averaged velocity $\langle v \rangle$ of the particle) versus the switching rate $\nu$ of trichotomous noise. The result of our precise numerical simulation of the system described by eq. (\ref{eq:model}) is compared with the analytical expression obtained by Mankin \emph{et al.} in \cite{mankin2000}, see eq. (27) and Fig. 1 therein. Parameters are: $q = 0.3$, $a_0 = 9$, $d = 0.25$, $F = 0$.}
	\label{fig2}
\end{figure}

The transport observable of foremost interest is  the stationary average velocity $\langle v \rangle$ which can be obtained from the formula 
\begin{equation}
	J  = \langle v \rangle = \lim_{t \to \infty} \frac{\langle x(t) \rangle}{t},
\end{equation}
where $J$ is a stationary probability current and $\langle \cdot \rangle$ denotes the average over all realizations of  trichotomous noise as well as over initial values of the process $x(t)$. From the symmetry considerations it follows that $F \to -F$ implies $\langle v \rangle \to -\langle v \rangle$ and  in particular $\langle v \rangle = 0$ for $F = 0$. 
For sufficiently small values of $F$ a linear response regime holds true and then 
\begin{equation}
	\langle v \rangle = \mu F.
\end{equation} 
In the normal transport regime the mobility coefficient $\mu$ is positive $\mu > 0$, whereas it is negative $\mu < 0$ when ANM occurs. In Ref. \cite{machura2007} the minimal model has been proposed for ANM in symmetric one-dimensional potentials: (i) the symmetric time-periodic driving of zero-mean that drives the system into a non-equilibrium state and (ii) the inertial term has to be included in the Langevin equation. It plays a \emph{crucial role} for this anomalous transport feature. In turn in Ref. \cite{spiechowicz2013,spiechowicz2014pre} it is shown that the constant force $F$ can be replaced by biased random noise of  non-zero mean value and then the counterpart of ANM can also be detected. 

The authors of Ref. \cite{chen2015} claim that ANM is also possible in a simpler case, i.e. in the overdamped regime where inertial term can be neglected. In order to verify this important statement  we resort to a precise numerical simulation of trichotomous noise driven Langevin dynamics determined by Eq. (\ref{eq:model}). All numerical calculations were done by use of a CUDA environment implemented on a modern desktop GPU. This proceeding allowed for a speed-up of a factor of the order $10^3$ as compared to a common present-day CPU method. For up-to-date review of this scheme we refer the reader to Ref. \cite{spiechowicz2015cpc}. We have performed several basic tests to check accuracy of our numerical codes for generation of trichotomous noise $\xi(t)$ and simulation of Langevin equation with the random force $\xi(t)$. In Fig. \ref{fig1} we compare the direct numerical results with the analytical expression for the correlation function $C(t)$ of noise $\xi(t)$. Other statistical characteristics of  noise have also been tested. All are in perfect agreement with the theoretical predictions. Moreover, we have carried out simulations of the system described by Eq. (\ref{eq:model}) for the particular case $F=0$ and compared numerical data with exact analytical results obtained by Mankin \emph{et al.} in \cite{mankin2000}, see eq. (27) and Fig. 1 therein. Again perfect agreement is noted, for details see Fig. \ref{fig2}.  

\begin{figure}[t]
	\centering
	\includegraphics[width=0.45\linewidth]{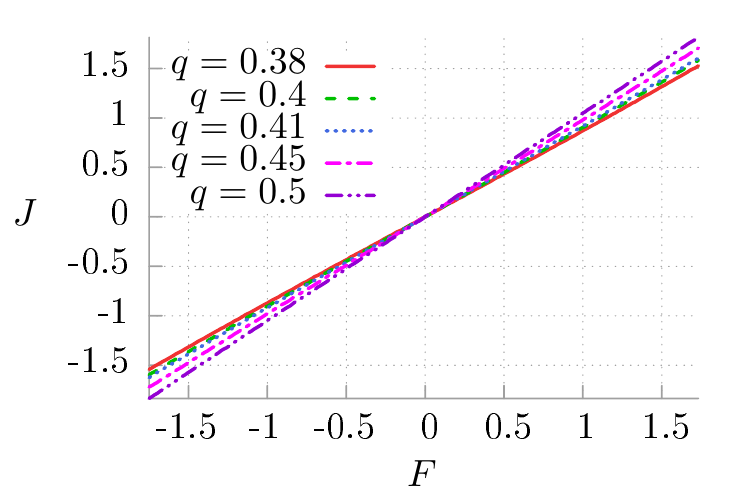}
	\caption{The probability current $J$ (which is equal to the stationary averaged velocity $\langle v \rangle$ of the particle) versus the constant force $F$ for selected values of the stationary probability $q$ of trichotomous noise. It is a precise numerical replica of Fig. 1 presented in \cite{chen2015}. According to the present state of the art about ANM it does not occur in the system described by Eq. (\ref{eq:model}). Parameters are $a_0 = 9$, $\nu = 10$ and $d = 0.5$.}
	\label{fig3}
\end{figure}
In Ref. \cite{chen2015}, the authors apply  the expression for the steady state probability current $J$ (which is equal to the stationary averaged velocity $\langle v \rangle$) which is obtained from the exact solution of the master (Fokker-Planck type) equation corresponding to Eq. (\ref{eq:model}) and derived in Ref. \cite{mankin2000}. However, the derivation presented there is valid for the case $F=0$ and under the assumption that  the potential $V(x)$ is differentiable at every point, so the piecewise linear potential has to be considered as a limit case of the smooth potential additionally fulfilling the conditions: 
\begin{equation}
	h(0) = 0 \quad \mbox{and} \quad h(d)  = 0,
\end{equation} 
where the conservative force is  $h(x) = -V'(x)$.  Surely, such  conditions do not hold when the  constant force $F$ acts on the particle since then the conservative force is $h(x) = -U'(x)$ with $U(x) = V(x) - Fx$ and generally
\begin{equation}
	h(0) \neq 0 \quad \mbox{and} \quad h(d) \neq 0.
\end{equation}
Therefore we conclude that the formula for the steady state probability current $J$ introduced in \cite{mankin2000} is no more valid when the bias $F$ is present in the system. Consequently, the use of that expression leads to incorrect results as it was the case in \cite{chen2015}. Instead of such proceeding, in Fig. \ref{fig3} we reproduce the parameter regime analysed in Fig. 1 of Ref. \cite{chen2015} with a direct precise numerical simulation of the dynamics described by Eq. (\ref{eq:model}). It follows that the mobility coefficient $\mu > 0$ and ANM does not occur in this overdamped system.

In summary, our results are fully consistent with the present state of the art about ANM. In a simple one-dimensional setup the \emph{inertial term} is essential to observe ANM. However, this statement is no longer true in a system of at least two coupled overdamped particles as then an interaction induced negative mobility phenomenon may occur \cite{januszewski2011}, 
also for Brownian transport containing a complex
topology \cite{EicRei2002a,EicRei2002b} and in some stylized,
multi-state models with state-dependent noise \cite{Haljas}. 

\section*{Acknowledgments}
This work was supported in part by the MNiSW program ”Diamond Grant” (J.S.).

\end{document}